\documentclass[12pt,a4paper]{article}
\usepackage{hyperref,amsmath,amssymb,array,framed}

\newcommand \vev [1] {\langle{#1}\rangle}

\DeclareMathOperator{\Conf}{Conf}
\DeclareMathOperator{\Gr}{Gr}
\DeclareMathOperator{\Li}{Li}

\begin{document}

\thispagestyle{empty}

\null\vskip-12pt \hfill CERN-PH-TH-2014-256\\
\null\vskip-12pt \hfill LAPTH-232/14 \\

\vskip2.2truecm
\begin{center}
\vskip 0.2truecm {\Large\bf
{\Large A Symbol of Uniqueness:\\ ~ \\ The Cluster Bootstrap for the 3-Loop MHV Heptagon}
}\\
\vskip 1truecm
{\bf J.~M. Drummond${}^{1,2,3}$, G. Papathanasiou${}^{3}$ and M. Spradlin${}^{4}$\\
}

\vskip 0.4truecm
{\it
${}^{1}$ School of Physics \& Astronomy, University of Southampton\\
Highfield, Southampton, SO17 1BJ, United Kingdom\\
\vskip .2truecm                        }
\vskip .2truecm
{\it
${}^{2}$ Theory Division, Physics Department, CERN\\
CH-1211 Geneva 23, Switzerland\\
\vskip .2truecm                        }
\vskip .2truecm
{\it
${}^{3}$ LAPTh, CNRS, Universit\'{e} de Savoie\\
F-74941 Annecy-le-Vieux Cedex, France\\
\vskip .2truecm                        }
\vskip .2truecm
{\it
${}^{4}$ Department of Physics, Brown University\\
Providence, RI 02912, USA\\
\vskip .2truecm                        }
\end{center}

\vskip 1truecm
\centerline{\bf Abstract}

Seven-particle scattering amplitudes in planar super-Yang-Mills theory are believed to belong to a special class of generalised polylogarithm functions called heptagon functions. These are functions with physical branch cuts whose symbols may be written in terms of the 42 cluster $\mathcal{A}$-coordinates on $\Gr(4,7)$. Motivated by the success of the hexagon bootstrap programme for constructing six-particle amplitudes we initiate the systematic study of the symbols of heptagon functions. We find that there is exactly one such symbol of weight six which satisfies the MHV last-entry condition and is finite in the $7 \parallel 6$ collinear limit. This unique symbol is both dihedral and parity-symmetric, and remarkably its collinear limit is exactly the symbol of the three-loop six-particle MHV amplitude, although none of these properties were assumed a priori. It must therefore be the symbol of the three-loop seven-particle MHV amplitude. The simplicity of its construction suggests that the $n$-gon bootstrap may be surprisingly powerful for $n>6$.

\newpage
\setcounter{page}{1}\setcounter{footnote}{0}
\tableofcontents

\section{Introduction}

A dream goal of the analytic S-matrix programme is to be able to construct expressions for the scattering amplitudes of a quantum field theory based on a few physical principles and a thorough knowledge of the analytic structure. In this work we are able to tie together recent advances in determining amplitudes by an analytic ``bootstrap'' procedure with discoveries about general classes of analytic functions which appear to play a central role. The theory we will study is the planar $\mathcal{N}=4$ supersymmetric gauge (SYM) theory in four dimensions~\cite{Brink:1976bc}, where the greatest advances have been made in explicitly determining the scattering amplitudes.

The analytic structure of the S-matrices of general quantum field theories are notoriously complicated~\cite{ELOP}. For the planar $\mathcal{N}=4$ super Yang-Mills theory however, several simplifying features come into play which reduce the complexity sufficiently to allow conjectures to be made about which classes of functions describe the scattering amplitudes, at least in the simplest cases. The duality with Wilson loops~\cite{Alday:2007hr,Drummond:2007aua,Brandhuber:2007yx,Drummond:2007cf,Drummond:2007au,Bern:2008ap,Drummond:2008aq} and the associated dual conformal symmetry~\cite{Drummond:2007au,Drummond:2006rz,Bern:2006ew,Bern:2007ct,Alday:2007he} of the planar theory mean that only amplitudes with six or more external legs are non-trivial. Moreover, for the six-particle case, the same amplitude/Wilson loop duality has allowed some explicit results~\cite{DelDuca:2010zg} and beautifully simple expressions~\cite{Goncharov:2010jf} to be obtained.

All of the above developments have led to an analytic bootstrap programme, so far focused on the six-particle (``hexagon'') case~\cite{Dixon:2011pw,Dixon:2011nj,Dixon:2013eka,Dixon:2014voa,Dixon:2014xca,Dixon:2014iba}. The idea of the hexagon bootstrap programme is to declare that, order by order in perturbation theory, six-particle amplitudes can be determined in terms of a particular class of multiple polylogarithms called hexagon functions. Hexagon functions are polylogarithms associated to a natural nine-letter alphabet of singularities which can be identified with the nine multiplicatively independent cross-ratios one can form from six points in $\mathbb{CP}^1$. In addition, hexagon functions obey conditions on the locations of branch cuts, encoding the fact that amplitudes can have discontinuities only in certain kinematical regions.

In the case of maximally helicity-violating (MHV) amplitudes, the relevant piece (obtained by subtracting particular universal infrared-divergent terms~\cite{Anastasiou:2003kj,Bern:2005iz}) which is not fixed by dual conformal symmetry is called the remainder function. At $L$ loops the $n$-particle remainder function $R_n^{(L)}$ should be a polylogarithm of weight $2L$ obeying additional criteria coming from various physical constraints. One constraint is that the remainder function should be fully dihedrally invariant, that is invariant under cyclic permutations $i \to i+1$ and flips $i \to n+1-i$ of its particle labels. This is essentially because supersymmetry dictates that the MHV amplitudes are given by an overall supersymmetric~\cite{Nair:1988bq,Parke:1986gb} Nair-Parke-Taylor factor at tree-level which then receives multiplicative quantum corrections. In addition the remainder function should approach $R_n^{(L)}\to R_{n-1}^{(L)}$ smoothly (i.e.~with power suppressed corrections) in the limit where the momenta of two colour-adjacent particles become collinear.

Moreover, the relation to Wilson loops means that the remainder function must obey constraints on its discontinuities~\cite{Alday:2010ku,Gaiotto:2010fk,Gaiotto:2011dt,Sever:2011pc} and on the power-suppressed corrections~\cite{Basso:2013vsa,Basso:2013aha,Basso:2014koa,Basso:2014nra,Belitsky:2014sla,Belitsky:2014lta,Basso:2014hfa} in the collinear limit coming from an operator product expansion (OPE) for light-like Wilson loops. The latter expansion is governed by the dynamics of an integrable colour-electric flux-tube, which in particular gives rise to all-loop integral formulas for individual power-suppressed terms. The success in systematically evaluating these in the weak coupling expansion for the first few terms~\cite{Papathanasiou:2013uoa,Papathanasiou:2014yva} gives hope that it may be even possible to resum the OPE (see~\cite{dp} for a first step in this direction) to obtain full amplitudes. More generally, it is expected that the integrability of the theory~\cite{Beisert:2010jr} will play an instrumental role in determining its S-matrix, and apart from the collinear limit it has also led to all-loop expressions in the multi-Regge limit~\cite{Basso:2014pla}, another kinematical regime that has provided significant information on the remainder function~\cite{Bartels:2008ce,Bartels:2008sc,Lipatov:2010qg,Lipatov:2010ad,Bartels:2010tx,Fadin:2011we,Prygarin:2011gd,Lipatov:2012gk,Dixon:2012yy, Bartels:2011ge,Bartels:2013jna,Bartels:2014jya}.

Finally, the extension of dual conformal symmetry to dual superconformal symmetry~\cite{Drummond:2008vq}, is expressed via the super-Wilson-loop correspondence~\cite{Mason:2010yk,CaronHuot:2010ek,Adamo:2011pv} in terms of recursive differential equations~\cite{Bullimore:2011kg,CaronHuot:2011kk} which imply certain universal constraints on the total derivative of the remainder function. Including the original superconformal symmetry of the amplitudes, or equivalently invoking parity for the amplitudes, extends dual superconformal symmetry to its Yangian~\cite{Drummond:2009fd,Drummond:2010qh}, leading to further recursive equations, relevant for determining non-MHV amplitudes.

The hexagon bootstrap programme has yielded explicit expressions up to four loops for the MHV amplitudes~\cite{Dixon:2011pw,Dixon:2013eka,Dixon:2014voa} and three loops in the NMHV case~\cite{Dixon:2011nj,Dixon:2014iba}. For higher multiplicities, explicit results in SYM theory so far have been confined (in general kinematics) to two loops~\cite{DelDuca:2009au,DelDuca:2010zp,Heslop:2010kq,CaronHuot:2011ky}. With explicit two-loop results to hand, an important structural observation has been made~\cite{Golden:2013xva}: the results are so far consistent with the conjecture that the relevant classes of functions are given by multiple polylogarithms whose singularities are dictated by a sequence of cluster algebras~\cite{1021.16017,1054.17024}. This observation will be of central importance here because it will allow us to generalise the bootstrap programme to higher multiplicities, beginning with seven-particle (``heptagon'') MHV amplitudes.

While one might expect that the bootstrap for heptagons would be a similar but more involved version of the bootstrap for hexagons, we are in fact led to a surprising and counterintuitive result. Up to three loops, to determine the symbol of the MHV remainder function, we need only construct the symbols of heptagon functions obeying the differential constraint coming from dual superconformal symmetry, and then demand that we have a linear combination of them which is finite in the collinear limit. Dihedral symmetry follows for free and no information coming from the OPE expansion of Wilson loops or the Regge limit of amplitudes is required at all. Moreover, the hexagon remainder can be obtained for free by taking the collinear limit of the heptagon remainder function. In this sense the heptagon bootstrap provides a conceptually more powerful framework for constructing even the hexagon amplitudes!

The plan of this paper is as follows. In section~\ref{sec:heptagonfunctions} we review the basic details needed to motivate the definition of heptagon functions. In section~\ref{sec:mhvconstraints} we review a few of the simplest general properties of MHV amplitudes in SYM theory, which in the bootstrap programme are applied as constraints on the space of heptagon functions. Section~\ref{sec:methods} contains a discussion of algorithms for imposing the constraint of integrability, which is by far the most significant computational challenge in applying the bootstrap. The expert reader may wish to jump directly (after taking a peek at the heptagon alphabet shown in eq.~(\ref{eq:heptagonletters})) to section~\ref{sec:heptagonproperties}, where our main results are discussed and summarised in Table~1. In section~\ref{sec:speculations} we attempt to formulate some explanation for why the heptagon (and higher-$n$) bootstrap is unexpectedly powerful.

Attached to the arXiv submission of this paper the reader may find data files containing: (1) the heptagon symbol alphabet shown in eq.~(\ref{eq:heptagonletters}), (2) the symbol of the remainder function $R_7^{(3)}$, and (3) the symbol of the other irreducible weight-6 heptagon function which satisfies the MHV last-entry condition (see subsection~\ref{subsec:unique}).

\section{Heptagon Functions}\label{sec:heptagonfunctions}

In this section we review some basic facts about generalised polylogarithms and symbols, leading up to our definition of heptagon functions which mirrors that of the hexagon functions studied in~\cite{Dixon:2011pw,Dixon:2011nj,Dixon:2013eka,Dixon:2014voa,Dixon:2014xca}.

\subsection{Symbols}

Precise definitions and additional details may be found in~\cite{Gonch3,Gonch2,FBThesis,Gonch} (see also~\cite{Duhr:2011zq} for a review), but here it is sufficient to recall the recursive definition according to which $f_k$ is called a generalised polylogarithm function of weight (or transcendentality) $k$ if its total differential may be written as a finite linear combination
\begin{equation}
\label{eq:dfk}
df_k = \sum_\alpha f_{k-1}^{(\alpha)}\, d \log \phi_\alpha
\end{equation}
over some set of $\phi_\alpha$, where the coefficients $f_{k-1}^{(\alpha)}$ are functions of weight $k-1$. Functions of weight 1 are defined to be finite linear combinations (with rational coefficients) of $\log \phi_\alpha$. By applying total derivatives $d$ to each of the coefficient functions $f_{k-1}^{(\alpha)}$ and using property~(\ref{eq:dfk}) recursively we arrive at a collection of rational numbers $f_0^{(\alpha_1,\alpha_2,\ldots,\alpha_k)}$ characterising the original function $f_k$. The symbol $\mathcal{S}(f_k)$ encapsulates this data via the definition
\begin{equation}
\label{eq:Sfkdef}
\mathcal{S}(f_k) = \sum_{\alpha_1,\ldots,\alpha_k} f_0^{(\alpha_1,\alpha_2,\ldots,\alpha_k)} \, (\phi_{\alpha_1} \otimes \cdots \otimes \phi_{\alpha_k})\,.
\end{equation}
Since $\log \phi_1 \phi_2 = \log \phi_1 + \log \phi_2$, it is evident that the consistency of eqs.~(\ref{eq:dfk}) and~(\ref{eq:Sfkdef}) requires symbols to satisfy
\begin{equation}
\label{eq:symbolidentity}
(\cdots \otimes \phi_1 \phi_2 \otimes \cdots) = (\cdots \otimes \phi_1 \otimes \cdots) +
(\cdots \otimes \phi_2 \otimes \cdots)\,.
\end{equation}
Moreover,
\begin{equation}
(\cdots \otimes c \otimes \cdots) = 0
\end{equation}
for any numerical constant $c$, since $d\log c = 0$. The collection of $\phi_\alpha$ which appear in the symbol of a given function is called its {\bf symbol alphabet}. A symbol alphabet is never uniquely defined because one can use the identity~(\ref{eq:symbolidentity}) to write symbols in various ways. Two alphabets $\{ \phi_\alpha \}$, $\{ \phi_\alpha'\}$ are considered equivalent if there exists a linear transformation
\begin{equation}
\label{eq:changeofbasis}
\log \phi_\alpha = \sum_\beta M_{\alpha \beta} \, \log \phi'_\beta
\end{equation}
given by an invertible matrix $M$ whose entries are rational numbers.

We consider the set of polylogarithm functions of weight $k$ as a vector space over the rational numbers. Moreover, since the product of two functions of weights $k_1$ and $k_2$ is a function of weight $k_1 + k_2$, they constitute a graded algebra. The irreducible elements of this algebra are functions which cannot be written as products of lower-weight functions.

The algebra of generalised polylogarithm functions admits a coproduct $\Delta$ compatible with multiplication, rendering it a Hopf algebra~\cite{Gonch2} (see also~\cite{Duhr:2012fh} for a review aimed at physicists). Moreover there is a cobracket $\delta$ which squares to zero (when acting on the quotient space of all functions modulo products of lower-weight functions), giving this algebra the structure of a Lie co-algebra. We will make no direct use of these higher mathematical structures in the present paper, but these tools have been very useful in elucidating the structure of two-loop MHV amplitudes~\cite{Golden:2013xva,Golden:2013lha,Golden:2014xqa,Golden:2014xqf} and in particular the relation between their coproducts and the cluster Poisson structure~\cite{FG03b} on the kinematical space on which they are defined.

\subsection{Symbol Alphabets}

A fundamental assumption of the ``cluster bootstrap'' programme is that the symbol alphabet relevant for $n$-particle amplitudes in SYM theory consists of the special collection of functions called cluster $\mathcal{A}$-coordinates on the kinematical configuration space $\Conf_n(\mathbb{P}^3) = \Gr(4,n)/(\mathbb{C}^*)^{n-1}$. We describe the space as kinematical because the $n$ points $Z_i$ (known as momentum twistors~\cite{Hodges:2009hk}) in $\mathbb{P}^3$ define, after choosing a preferred bitwistor $I \in \mathbb{P}^3 \wedge \mathbb{P}^3$, a light-like polygonal contour in Minkowski space-time via
\begin{equation}
x_i \sim \frac{Z_{i-1}\wedge Z_i}{\langle Z_{i-1} Z_i I\rangle}\,,
\end{equation}
where $\langle Z_{i} Z_j Z_{k} Z_l \rangle=\det(Z_i Z_j Z_k Z_l)$. The particle momenta can be identified with the null separations of neighbouring points,
\begin{equation}
p_i = x_{i+1} - x_i\,,
\end{equation}
and the kinematical Mandelstam variables can be identified with the non-zero separations and hence related to the momentum twistors,
\begin{equation}
(p_{i} + p_{i+1} + \ldots p_{j-1})^2 = (x_i - x_j)^2 = \frac{\langle Z_{i-1} Z_i Z_{j-1} Z_j \rangle}{\langle Z_{i-1} Z_i I \rangle \langle Z_{j-1} Z_j I \rangle}\,.
\end{equation}

There is a vast mathematical literature on cluster algebras; we refer the reader to~\cite{Golden:2013xva} for an introduction focused on amplitudes. The cluster $\mathcal{A}$-coordinates relevant to $n$-particle amplitudes consist of the Pl\"ucker coordinates $\vev{ijkl} \equiv \langle Z_i Z_j Z_k Z_l\rangle$ which can be formed from the momentum twistors $Z_i$ specifying the kinematics of the scattering particles, together with certain very particular homogeneous polynomials in Pl\"ucker coordinates which can be systematically constructed via an algorithm known as mutation.

The fact that amplitudes in SYM theory depend on the individual $Z_i$ only through the (projective) $SL(4)$ invariants $\vev{ijkl}$ is a consequence of dual conformal symmetry. The individual four-brackets are not invariant under projective transformations of the homogeneous coordinates $Z_i$ on $\mathbb{P}^3$, so they must always appear in projectively invariant ratios.

The case $n=6$ is the simplest, since mutation does not generate any $\mathcal{A}$-coordinates beyond the standard Pl\"ucker coordinates on $\Gr(4,6)$. From the 15 individual four-brackets we can form 9 invariant ratios, for example
\begin{align}
u &= \frac{\vev{6123}\vev{3456}}{\vev{6134}\vev{2356}}\,,
&
v &= \frac{\vev{1234}\vev{4561}}{\vev{1245}\vev{3461}}\,,
&
w &= \frac{\vev{2345}\vev{5612}}{\vev{2356}\vev{4512}}\,,
\nonumber \\
1{-}u &= \frac{\vev{5613}\vev{6234}}{\vev{6134}\vev{2356}}\,,
&
1{-}v &= \frac{\vev{6124}\vev{1345}}{\vev{1245}\vev{3461}}\,,
&
1{-}w &= \frac{\vev{1235}\vev{2456}}{\vev{2356}\vev{4512}}\,,
\label{eq:hexagonletters}
\\
y_u &= \frac{\vev{1345}\vev{2456}\vev{1236}}{\vev{1235}\vev{3456}\vev{1246}}\,,
&
y_v &= \frac{\vev{1235}\vev{2346}\vev{1456}}{\vev{1234}\vev{2456}\vev{1356}}\,,
&
y_w &= \frac{\vev{2345}\vev{1356}\vev{1246}}{\vev{1345}\vev{2346}\vev{1256}}\,.
\nonumber
\end{align}
This particular choice of basis has been widely used in the literature, but as mentioned above any multiplicatively transformed set of ratios would serve just as well, if one is interested in working only at the level of symbols. The so-called hexagon bootstrap is predicated on the assumption that all $L$-loop six-particle amplitudes (both MHV and non-MHV) in SYM theory are generalised polylogarithm functions of weight $k=2L$ whose symbols can be written in terms of the nine-letter alphabet shown in eq.~(\ref{eq:hexagonletters}). This hypothesis has been successfully tested for the MHV remainder function through four loops~\cite{Dixon:2011pw,Dixon:2013eka,Dixon:2014voa} and the NMHV ratio function through three loops~\cite{Dixon:2011nj,Dixon:2014iba}. Further support for its validity comes from a particular ``dlog'' representation of the all-loop integrand~\cite{ArkaniHamed:2012nw}, as well as the all-loop basis of harmonic polylogarithms found for the first few orders of these amplitudes in an expansion around the collinear limit~\cite{Papathanasiou:2013uoa,Papathanasiou:2014yva}.

For $n>7$ the cluster algebra associated to $\Conf_n(\mathbb{P}^3)$ has infinitely many $\mathcal{A}$-coordinates. This is not necessarily an obstacle to the cluster bootstrap programme as long as only a finite number of them appear at any finite order in perturbation theory. For example, it is known~\cite{CaronHuot:2011ky} that the two-loop $n$-particle MHV remainder function is written in terms of a symbol alphabet of precisely $\frac{3}{2} n (n-5)^2$ (projectively invariant) letters. It would be very interesting to determine whether, for example, the symbol of the three-loop eight-particle MHV remainder function may be written in terms of the same 108 letters which appear already at two loops, or whether it requires more exotic cluster $\mathcal{A}$-coordinates. This amplitude has been evaluated in in two-dimensional kinematics~\cite{Caron-Huot:2013vda}, but unfortunately this limit appears to be insufficient to decide the question.

In this paper we focus on the $n=7$ Goldilocks zone, where the number of $\mathcal{A}$-coordinates is still finite, but in addition to the Pl\"ucker coordinates $\vev{ijkl}$ there are 14 $\mathcal{A}$-coordinates which are bilinears of the form
\begin{equation}
\vev{a(bc)(de)(fg)} \equiv
\vev{abde} \vev{acfg} - \vev{abfg} \vev{acde}\,.
\end{equation}
This notation emphasises the antisymmetry under exchange of any pair of indices inside parentheses, as well as antisymmetry under the exchange of the pairs amongst each other. In the mathematical literature on cluster algebras the $n$ cyclic Pl\"ucker coordinates $\vev{i\,i{+}1\,i{+}2\,i{+}3}$ are usually treated differently and are sometimes called ``coefficients'' instead of ``coordinates''. With this terminology, there are precisely 42 cluster $\mathcal{A}$-coordinates for the case $n=7$, given by
\begin{equation}
\label{eq:sixAs}
\vev{2367}\,, \quad
\vev{2567}\,, \quad
\vev{2347}\,, \quad
\vev{2457}\,, \quad
\vev{1(23)(45)(67)}\,, \quad
\text{and~}
\vev{1(34)(56)(72)}\,,
\end{equation}
together with their images under cyclic transformations $Z_i \to Z_{i+1}$.

Projectively invariant ratios can be formed by dressing each of these 42 coordinates with suitable powers of the $\vev{i\,i{+}1\,i{+}2\,i{+}3}$ Pl\"ucker coordinates, as it is always possible to construct products of the latter with helicity weight at a single point, and combine them so as to cancel the excess weight of the points appearing in (\ref{eq:sixAs}). We have found a convenient choice to be
\begin{align}
\label{eq:heptagonletters}
a_{11} &=
\frac{\vev{1234} \vev{1567} \vev{2367}}{\vev{1237} \vev{1267} \vev{3456}}\,,
&
a_{41} &=
\frac{\vev{2457} \vev{3456}}{\vev{2345} \vev{4567}}\,,
\nonumber
\\
a_{21} &=
\frac{\vev{1234} \vev{2567}}{\vev{1267} \vev{2345}}\,,
&
a_{51} &=
\frac{\vev{1(23)(45)(67)}}{\vev{1234} \vev{1567}}\,,
\\
a_{31} &=
\frac{\vev{1567} \vev{2347}}{\vev{1237} \vev{4567}}\,,
&
a_{61} &=
\frac{\vev{1(34)(56)(72)}}{\vev{1234} \vev{1567}}\,,
\nonumber
\end{align}
together with $a_{ij}$ obtained from $a_{i1}$ by cyclically relabeling $Z_m \to Z_{m + j - 1}$. While the $a_{ij}$ are multiplicatively independent, they are of course not algebraically independent: the dimension of $\Conf_7(\mathbb{P}^3)$ is only six, so one could choose to parameterise all 42 of the $a_{ij}$ in terms of just 6 free variables if needed.

As noted in eq.~(\ref{eq:changeofbasis}) the choice of symbol alphabet is not unique or canonical. In contrast, it has been noted~\cite{Golden:2013xva,Golden:2014pua} that the coproducts of two-loop MHV remainder functions $R_n^{(2)}$ involve only preferred cross-ratios known as cluster $\mathcal{X}$-coordinates on $\Conf_n(\mathbb{P}^3)$. None of the $a_{ij}$ in eq.~(\ref{eq:heptagonletters}) are cluster $\mathcal{X}$-coordinates for $n=7$. The latter, which have been tabulated in section 7.3 of~\cite{Golden:2013xva}, may be expressed as products of powers of the former. It would be interesting to understand if there is a connection between coproducts and $\mathcal{X}$-coordinates beyond two loops, or for non-MHV amplitudes.

It is also interesting to note that only 14 out of the 105 possible distinct $\vev{a(bc)(de)(fg)}$ objects appear in eq.~(\ref{eq:sixAs}). This is indicative of a qualitative difference between the cases $n=6$ and $n>6$. For $n=6$ the set of $\mathcal{A}$-coordinates, as a whole, is invariant (up to overall signs, which are never a concern inside symbols) under arbitrary permutations of the particle labels, not just under cyclic permutations. However for $n>6$, a non-cyclic permutation would actually change the symbol alphabet. For example, switching $1 \leftrightarrow 4$ would have no substantive effect on eq.~(\ref{eq:hexagonletters}) (it would rearrange the letters to an equivalent basis), but it would completely change the heptagon basis~(\ref{eq:heptagonletters}) by introducing genuinely new letters which are not cyclic rotations of those in eq.~(\ref{eq:sixAs}).

This dependence of the symbol alphabet on the choice of dihedral structure, i.e.~on a particular ordering of the particles, is in fact natural. We recall that when we refer to ``amplitude" we really mean the colour-ordered partial amplitude $\mathcal{A}(1,\ldots,n)$ which produces the full amplitude upon summation over non-cyclic permutations $\sigma$,
\begin{equation}
\mathcal{A}_{\rm full} = \sum_\sigma {\rm Tr} (T^{a_{\sigma(1)}} \ldots T^{a_{\sigma(n)}})\, \mathcal{A}(\sigma(1),\ldots,\sigma(n))\,.
\end{equation}
Thus, while for six particles, each colour-ordered partial amplitude is described by one and the same class of polylogarithms, the general case requires different classes of polylogarithms for different colour-ordered partial amplitudes.

The {\bf heptagon bootstrap} which we initiate in this paper is based on the hypothesis that all $L$-loop seven-particle amplitudes (whether MHV or non-MHV) are generalised polylogarithm functions of weight $k=2L$ whose symbols can be written in terms of the 42-letter alphabet shown in eq.~(\ref{eq:heptagonletters}).

\subsection{Integrable Words}

Given a random symbol $\mathcal{S}$ of weight $k>1$, there does not in general exist any function whose symbol is $\mathcal{S}$. A symbol of the form~(\ref{eq:Sfkdef}) is said to be {\bf integrable}, (or, to be an {\bf integrable word}) if it satisfies
\begin{equation}
\label{eq:integrability}
\sum_{\alpha_1,\ldots,\alpha_k} f_0^{(\alpha_1,\alpha_2,\ldots,\alpha_k)}
\underbrace{(\phi_{\alpha_1} \otimes \cdots \otimes \phi_{\alpha_k})}_{\text{omitting $\phi_{\alpha_{j}} \otimes
\phi_{\alpha_{j{+}1}}$}}
\ d \log \phi_{\alpha_j} \wedge d \log \phi_{\alpha_{j{+}1}} = 0 \qquad \forall j \in
\{1,\ldots,k-1\}\,.
\end{equation}
These are necessary and sufficient conditions for a function $f_k$ with symbol $\mathcal{S}$ to exist.

There are $42^k$ distinct symbols of weight $k$ which can be written in the 42-letter symbol alphabet shown in eq.~(\ref{eq:heptagonletters}), but only certain linear combinations of these satisfy the integrability conditions~(\ref{eq:integrability}). Determining these linear combinations is, in general, a computationally difficult problem which we discuss in detail in section~\ref{sec:methods} below. For weight $k=1,2,3$, only $42, 1035, 19536$ linear combinations of the $42^k = 42, 1764, 74088$ available symbols are integrable, and hence correspond to actual functions.

It is relatively easy to tabulate these functions explicitly. We begin with the fact that any generalised polylogarithm function of weight 3 or less can be written in terms of the classical polylogarithm functions $\Li_k(x)$. Since
\begin{equation}
\mathcal{S}(\Li_k(x)) = - (1 - x) \otimes \underbrace{x \otimes \cdots \otimes x}_{\text{$k{-}1$ times}}
\end{equation}
we can allow the argument $x$ to be any product of powers of the $a_{ij}$ with the property that $1-x$ can also be expressed as a product of powers of $a_{ij}$'s. There are precisely 2310 distinct $x$'s of this type.

At weight 2 not all 2310 of the $\Li_2(x)$'s are independent since there are many identities for the $\Li_2$ function. These include $\Li_2(x) \approx - \Li_2(1/x) \approx - \Li_2(1 - x)$ (where $\approx$ means modulo products of functions of lower weight, i.e.~modulo $\mathcal{O}(\log^2)$ in this case), as well as the pentagon identity. It can be checked that only 132 out of the 2310 $\Li_2(x)$'s are linearly independent mod $\mathcal{O}(\log^2)$. Hence the vector space of irreducible weight-2 integrable words has dimension 132.

At weight 3 we have the identities $\Li_3(x) \approx \Li_3(1/x)$ and $\Li_3(x) + \Li_3(1-x) + \Li_3(1-1/x) \approx 0$, which leave $2310/3 = 770$ independent functions. There are also 22 linearly independent $D_4$ identities~\cite{Golden:2013xva}, so we see that there are precisely 748 linearly independent irreducible weight-3 integrable words.

Having determined that there are 42, 132, 748 irreducible functions at weight $k=1,2,3$, it is simple to consider all possible ways of taking products of lower-weight functions to count the total number of functions $42, 1035, 19536$ given above. Let us stress that here we have counted all functions written from the heptagon alphabet, but we will only give the name ``heptagon function'' to the subset satisfying an important analytic constraint to which we now turn our attention.

\subsection{Physical Singularities}

Most of the functions discussed in the previous section have no possible relevance to amplitudes. One simple criterion which eliminates many of them is locality, which imposes tight constraints on the analytic properties of any scattering amplitude. In particular, it is a basic consequence of locality that amplitudes may only have singularities when some intermediate particle goes on-shell. For planar colour-ordered amplitudes in massless theories this can only happen when some sum of cyclically adjacent momenta $p_i + p_{i+1} + \cdots + p_{j-1} = x_j - x_i$ becomes null. The Euclidean region, in which amplitudes must be free of branch points, corresponds to having all non-neighbouring separations $x_j - x_i$ space-like.

The effect of this branch cut condition on the symbol of a seven-particle amplitude is that only the seven $a_{1j}$ are allowed to appear in the first entry. This is because the singularities of generalised polylogarithm functions are encoded in the first entry of their symbols: specifically, a letter $\phi$ appearing in the first entry indicates that the corresponding function has branch points at $\phi=0$ and $\phi=\infty$. From eq.~(\ref{eq:heptagonletters}) we see that only the $a_{1j}$ are built out of dual conformal invariant cross-ratios which may be formed from the $(x_j - x_i)^2$; the other letters contain quantities which could not possibly cancel between different additive terms in a symbol since the $a_{ij}$ are multiplicatively independent. The restriction that only these cross-ratios may appear in the first entry is referred to as the {\bf first-entry condition}.

\subsection{Heptagon Functions}\label{subsec:heptagonfunctions}

Following the definition of hexagon functions given in~\cite{Dixon:2013eka}, we define a {\bf heptagon function} of weight $k$ to be a polylogarithm function of weight $k$ whose symbol may be written in the alphabet~(\ref{eq:heptagonletters}) and which is free of branch points in the Euclidean region. As discussed in the previous subsection, such functions have symbols in which only the letters $a_{1j}$ appear in the first entry.

We follow the standard convention of counting heptagon functions of a certain weight only modulo the addition of functions of lower weight (times numerical constants of the appropriate transcendental weight). Although in this paper we work entirely at the level of symbols, if we restrict to the definitions and conventions we have introduced so far, the counting of the heptagon functions and the counting of their symbols will coincide.

More generally however, it is important to note that when additional constraints are imposed, the number of heptagon functions satisfying them may be smaller than the number of their symbols. In particular, it can happen that a symbol which satisfies the first-entry condition and is well-defined in a collinear limit can be promoted to a function with physical branch cuts only by adding certain terms of lower weight, which may end up diverging in the collinear limit. An example of this phenomenon has already been seen at three loops in the MHV hexagon case~\cite{Dixon:2011pw,Dixon:2013eka}.

Bearing this caveat in mind, especially in light of the fact that we will be examining collinear limits in what follows, we will be careful to only interchange the terms ``heptagon function'' and ``symbol of heptagon function'' when the counting coincides, and otherwise employ the term {\bf heptagon symbol} to denote the latter in a more abbreviated fashion. Finally, it should be understood that we are really counting dimensions of vector spaces of symbols, not individual symbols, so when we say there is a unique symbol with certain properties, we mean unique up to an overall multiplicative factor.

Using the algorithms described in section~\ref{sec:methods} below, we have found that the dimension of the space of heptagon functions is 7, 42, 237, 1288, 6763 for $k=1,2,3,4,5$. These numbers, and the dimensions of various physically interesting subspaces, are tabulated in Table~1. It follows from this counting that the vector space of irreducible heptagon functions has dimension 7, 14, 55, 196, 708 for $k=1,2,3,4,5$.

\section{MHV Constraints}\label{sec:mhvconstraints}

We believe that all seven-particle amplitudes in SYM theory are heptagon functions as defined in the previous section. In this section we discuss some of the additional properties special to MHV amplitudes, which will be the focus of most of the remainder of the paper.

\subsection{The ${\bar Q}$ Equation}\label{subsec:qbar}

It has been argued in~\cite{CaronHuot:2011ky}, and subsequently shown to be a consequence of a proposed anomaly equation for the ${\bar Q}$ dual superconformal symmetry generators~\cite{CaronHuot:2011kk}, that the extended superconformal symmetry of SYM theory implies that the differential of any MHV amplitude can be written as a linear combination of $d \log \vev{i\,j{-}1\,j\,j{+}1}$. Evidently, from eqs.~(\ref{eq:dfk}) and~(\ref{eq:Sfkdef}), this implies that only the Pl\"ucker coordinates $\vev{i\,j{-}1\,j\,j{+}1}$ may appear in the last entry of the symbol of any MHV amplitude. This is called the {\bf last-entry condition}. For the case $n=7$, we see from eq.~(\ref{eq:heptagonletters}) that in our basis, only the 14 letters $a_{2j}$ and $a_{3j}$ may appear in the last entry of the symbol of the seven-particle MHV amplitude.

\subsection{The Collinear Limit}

MHV amplitudes have particularly simple behavior under collinear limits. It is baked into the definition of the BDS-subtracted $n$-particle $L$-loop MHV remainder function~\cite{Bern:1994zx,Bern:1997nh} that it should smoothly approach the corresponding $n{-}1$-particle function in any simple collinear limit:
\begin{equation}
\label{eq:collinear}
\lim_{i{+}1 \parallel i} R^{(L)}_n = R^{(L)}_{n-1}\,.
\end{equation}
Although we do not do so in the present paper, it would be interesting to also consider the constraints imposed by multi-collinear limits, under which MHV remainder functions have a more intricate behavior (see for example~\cite{Bern:2008ap,Anastasiou:2009kna}).

We can parameterise the $7 \parallel 6$ collinear limit as
\begin{equation}
\label{eq:Zcollinear}
Z_7 \to Z_6 + \epsilon \frac{\vev{1246}}{\vev{1245}} Z_5 +
\epsilon \tau \frac{\vev{2456}}{\vev{1245}} Z_1 +
\eta \frac{\vev{1456}}{\vev{1245}} Z_2\,,
\end{equation}
where the limit $\eta \to 0$ is taken first, followed by $\epsilon \to 0$, leaving the parameter $\tau$ fixed. The ratios of four-brackets in eq.~(\ref{eq:Zcollinear}) could be absorbed into $\epsilon$, $\eta$ and $\tau$, but these factors are useful for keeping track of twistor weight.

Under the replacement~(\ref{eq:Zcollinear}), the 42-letter heptagon symbol alphabet collapses into the 9-letter hexagon symbol alphabet shown in eq.~(\ref{eq:hexagonletters}) plus nine additional letters: the vanishing letters $\epsilon$ and $\eta$, as well as the seven finite letters
\begin{align}
\begin{split}
&\tau\,,
\cr
&1 + \tau\,,
\cr
&\vev{1235} \vev{1246} + \tau \vev{1236} \vev{1245}\,,
\cr
&\vev{1245} \vev{3456} + \tau \vev{1345} \vev{2456}\,,
\cr
&\vev{1246} \vev{2356} + \tau \vev{1236} \vev{2456}\,,
\cr
&\vev{1246} \vev{3456} + \tau \vev{1346} \vev{2456}\,,
\cr
&\vev{1235} \vev{1246} \vev{3456} + \tau \vev{1236} \vev{1345} \vev{2456}\,.
\label{eq:badletters}
\end{split}
\end{align}

A function has a well-defined $7 \parallel 6$ collinear limit only if its symbol is independent of all nine of these letters. We can parameterise other $i{+}1 \parallel i$ simple collinear limits by appropriately relabeling eq.~(\ref{eq:Zcollinear}) cyclically.

\subsection{Discrete Symmetries}

MHV amplitudes must satisfy several discrete symmetries. They are invariant under the $n$-particle dihedral group generated by cyclic transformations $Z_i \to Z_{i+1}$ as well as the flip (orientation reversal operation) $Z_i \to Z_{n+1 - i}$. These discrete symmetries act simply on the $a_{ij}$, taking each heptagon letter to some other, as may be read off from eq.~(\ref{eq:heptagonletters}).

A less trivial symmetry of MHV amplitudes is spacetime parity, which in momentum twistor space is generated by the involution
\begin{equation}
Z_i \to W_i \equiv \vev{*\,i{-}1\,i\,i{+}1}\,.
\end{equation}
This notation is meant to indicate that $W_i$ is a vector orthogonal to the hyperplane spanned by $Z_{i-1}$, $Z_i$ and $Z_{i+1}$. Under parity the letters $a_{1i}$ and $a_{6i}$ are invariant, while the others obey:
\begin{equation}
\label{eq:parity}
a_{21} \longleftrightarrow a_{37}\,, \qquad
a_{41} \longleftrightarrow a_{51}\,,
\end{equation}
and cyclically related transformations.

\section{Methods for Constructing Integrable Words}\label{sec:methods}

The problem of enumerating all integrable words of length $k$ written in a given alphabet is computationally challenging in general. An exception is when the symbol alphabet consists of cluster coordinates on $\Gr(2,n)$, corresponding to iterated integrals~\cite{Chen} on a Riemann sphere with $n$ marked points, in which case the functions may be explicitly enumerated~\cite{FBThesis}.

When the symbol alphabet is finite, as is the case for the 42-letter heptagon alphabet, at least it is a finite problem. Beginning with the vector space spanned by all $42^k$ (or fewer, if other conditions have been imposed) length-$k$ words, one needs simply to determine how many linear combinations satisfy the integrability constraints~(\ref{eq:integrability}). Since these are linear constraints, the problem of enumerating all integrable words is ultimately one of linear algebra: it is the problem of finding a basis for the kernel of the matrix of the integrability constraints.

The calculation may be organised in a couple of different ways, which have various advantages and disadvantages as we now discuss.

\subsection{A Stepwise Approach}\label{subsec:stepwiseapproach}

For low weights we can use a standard recursive method of iteratively constructing integrable words. First we make an ansatz for words of length $k$ by adjoining one extra letter in all possible ways to integrable words of length $k-1$ and then we directly impose integrability on the last two slots. For the final step of imposing integrability it is convenient to calculate once, and store the value of, all possible combinations $\omega_{\alpha \beta} = d \log \phi_\alpha \wedge d \log \phi_\beta$ as explicit two-forms expressed in some choice of variables.

In general the two-forms $\omega_{\alpha\beta}$ will be non-trivial functions of the $\phi$'s. The condition that eq.~(\ref{eq:integrability}) should vanish identically may be translated into a collection of linear equations by evaluating the equation at sufficiently many randomly selected points. The nullspace of this linear system is the vector space of integrable words. There is never any concern that an accidentally poor choice of random points may lead to an erroneously large nullspace (i.e., to mistakenly conclude that there are more integrable words than actually exist) because while solving eq.~(\ref{eq:integrability}) is difficult, it is completely straightforward to check whether or not any putative solution is valid.

\subsection{A Bootstrap}\label{subsec:bootstrap}

For higher weights we have found an alternative recursive method preferable. Let $\mathcal{A}$ denote the symbol alphabet, let $\mathcal{W}_k$ be the vector space of integrable words of length $k$ written in $\mathcal{A}$, and let $\{w_i^{(k)}\}$ be a basis for this space, where $i=1,\ldots,d_k = \dim(\mathcal{W}_k)$. Suppose that we have determined such a basis for all weights up to some value $k$. Then we can expand each basis element $w_i^{(k)}$ as a linear combination of words of the form $\mathcal{W}_{k-1} \otimes \mathcal{A}$ in order to make the last entry in each term explicit, i.e.
\begin{equation}
\label{eq:basisA}
w_l^{(k)} = \sum_{i=1}^{d_{k-1}} \sum_\alpha A_{li\alpha}^{(k)}
\ (w_i^{(k-1)} \otimes \phi_\alpha)
\end{equation}
for some rational coefficients $A$. Similarly, we can make the first entry in each term explicit by expanding in $\mathcal{A} \otimes \mathcal{W}_{k-1}$,
\begin{equation}
\label{eq:basisB}
w_m^{(k)} = \sum_{j=1}^{d_{k-1}} \sum_\alpha B_{mj\alpha}^{(k)}
\ (\phi_\alpha \otimes w_j^{(k-1)})\,.
\end{equation}
The $A$ and $B$ coefficients may be easily computed once bases for $\mathcal{W}_k$ and $\mathcal{W}_{k-1}$ are known.

Now let $1 < k_1, k_2 \le k$. We may then write an ansatz for words of length $k_1 + k_2 - 1$ as a linear combination of the form
\begin{equation}
\label{eq:C1w}
\sum_{l=1}^{d_{k_1}} \sum_{j=1}^{d_{k_2-1}} C_{lj}^{(1)}
\ ( w_l^{(k_1)} \otimes w_j^{(k_2-1)})
\end{equation}
for some rational coefficients $C^{(1)}$. This ansatz is manifestly integrable in the first $k_1$ entries, as well as in the last $k_2-1$ entries, so the coefficients $C^{(1)}$ are to be determined by imposing integrability only between entries $k_1$ and $k_1+1$. On the other hand we may write an alternative ansatz of the form
\begin{equation}
\label{eq:C2w}
\sum_{i=1}^{d_{k_1-1}} \sum_{m=1}^{d_{k_2}} C_{im}^{(2)}
\ ( w_i^{(k_1-1)} \otimes w_m^{(k_2)})
\end{equation}
where integrability is manifest everywhere except between entries $k_1-1$ and $k_1$.

Now any integrable word of length $k_1 + k_2-1$ must of course admit an expansion of both types simultaneously, so we can impose full integrability by equating the two forms of the ansatz. Using the basis decompositions~(\ref{eq:basisA}), (\ref{eq:basisB}) to expose the intermediate letter in slot $k_1$ lets us express the compatibility conditions as
\begin{equation}
\sum_{i,j,l,\alpha}
A^{(k_1)}_{li\alpha} C^{(1)}_{lj}
(w_i^{(k_1-1)} \otimes \phi_\alpha \otimes w_j^{(k_2-1)})
=
\sum_{i,j,m,\alpha}
B^{(k_2)}_{mj\alpha} C^{(2)}_{im}
(w_i^{(k_1-1)} \otimes \phi_\alpha \otimes w_j^{(k_2-1)})\,.
\end{equation}

We may express these equations more simply in matrix form: we have a $d_{k_1} \times d_{k_2-1}$ matrix $C^{(1)}$, a $d_{k_1-1} \times d_{k_2}$ matrix $C^{(2)}$, and, for each value $\alpha$ (i.e., for each letter in the alphabet) a $d_{k_1} \times d_{k_1-1}$ matrix $A^{(k_1)}_\alpha$ and a $d_{k_2} \times d_{k_2 -1}$ matrix $B^{(k_2)}_\alpha$, subject to the $d_{k_1-1} \times d_{k_2-1}$ matrix relations
\begin{equation}
\label{eq:ABintegrability}
(A^{(k_1)}_\alpha)^{\rm T} C^{(1)} = C^{(2)} B^{(k_2)}_\alpha \qquad \forall \alpha\,.
\end{equation}
Given the $A$'s and $B$'s constructed as described above, any solution $(C^{(1)}, C^{(2)})$ to this linear system determines an integrable word of length $k_1 + k_2 - 1$. Note that all of these matrices should have rational entries.

Although we have phrased it here in a general manner, this construction lets us impose the first- and/or last-entry conditions in a very straightforward way. For example, to impose the first-entry condition we simply restrict all of the above formulas from the $d_{k_1-1}, d_{k_1}$ dimensional spaces of all integrable words to the $\tilde{d}_{k_1-1}, \tilde{d}_{k-1}$ dimensional subspaces satisfying the first-entry condition. The same relation~(\ref{eq:ABintegrability}) holds, but with significantly smaller $A_\alpha^{(k_1)}$, $C^{(1)}$ and $C^{(2)}$ matrices. Imposing the last-entry restriction reduces the size of $C^{(1)}$ and $C^{(2)}$ further, and also the size of $B_\alpha^{(k_2)}$.

\subsection{Comparison of the Two Methods}\label{subsec:comparisontwomethods}

The only notable disadvantage of the new recursive approach is that the equations~(\ref{eq:ABintegrability}) to solve involve considerably more free variables. To illustrate this point, let us describe the construction of integrable words of length 6 satisfying both the first and last-entry conditions. The traditional approach of subsection~\ref{subsec:stepwiseapproach} involves an ansatz with $6763 \times 14 = 94682$ free parameters, corresponding (see Table~1) to the number of weight-5 heptagon functions, tensored with the 14 allowed last entries. The number of equations depends on the number of random kinematical points at which we evaluate eq.~(\ref{eq:integrability}), but should be at least comparable to the number of free variables.

In the new approach of subsection~\ref{subsec:bootstrap} we use eq.~(\ref{eq:ABintegrability}) with $k_1 = 4$ and $k_2 = 3$. From Table~1 we see that there are 237 (1288) heptagon functions at weight 3 (4), i.e.~integrable words of length 3 (4) satisfying the first-entry condition. Meanwhile one can check that there are 146 (1364) integrable words of length 2 (3) satisfying the last-entry condition. Therefore, applying eq.~(\ref{eq:ABintegrability}) to find the space of heptagon functions satisfying the last-entry condition requires solving $237\times 146\times 42$ equations for the $1288 \times 146$ matrix $C^{(1)}$ and the $237 \times 1364$ matrix $C^{(2)}$, i.e. a total of about one and a half million equations for over half a million free variables! (As discussed in the next section, the solution space of this system turns out, amazingly, to have dimension four.)

However, we have found this disadvantage to be more than compensated by two significant advantages. The first is that if the bases for the $\mathcal{W}_k$ are chosen with reasonable care, the matrices $A_\alpha$ and $B_\alpha$ can be made quite sparse. By solving the simplest conditions, namely those that force a single free parameter to vanish, or express one parameter in terms of exactly one another, we may quickly reduce the size of the linear system. This contrasts to the traditional iterative construction described at the beginning of this section, since the $d\log\phi_\alpha \wedge d\log \phi_\beta$ two-forms generally have no, or only a few, vanishing elements. For the length 6 linear system we discussed in the previous paragraphs, solving the equations of length 1 or 2 and then partially solving some more of the shorter constraints leads to 63557 equations for the remaining 15979 free variables, a significant reduction of the size of the linear system compared to the traditional approach.

The second great advantage of the method of subsection~\ref{subsec:bootstrap} is also illustrated in the length 6 example we have used: if bases are known for all weights less than or equal to some value $k$, we may immediately bootstrap directly to a basis of integrable words at weight $2k-1$, without having to recursively construct bases at weights $k+1,k+2,\ldots$ one step at a time, as with the traditional approach. (Because of this weight-skipping power of the bootstrap we had in fact found the symbol of the 3-loop MHV heptagon long before determining the total number of weight-5 heptagon functions.)

\subsection{Solving the Integrability Constraints}

Even with the improved method of subsection~\ref{subsec:bootstrap}, starting at weight 5 the size of the linear system encoding the integrability constraints grows to such extent that its solution becomes the most important computational challenge of the bootstrap programme. Let us now discuss the strategy we adopted for addressing this challenge, which required an efficiency beyond the capabilities of standard scientific software such as \texttt{Mathematica} or \texttt{Matlab}.

After (or even before) partially reducing the integrability constraints in the form of subsection~\ref{subsec:bootstrap} according to the discussion of subsection~\ref{subsec:comparisontwomethods}, we may bring them to a more standard form by grouping all elements of the matrices of unknown coefficients $C^{(1)},C^{(2)}$ into a column vector $X$, such that eq.~(\ref{eq:ABintegrability}) becomes
\begin{equation}\label{eq:AX=0}
A\cdot X=0\,.
\end{equation}
By virtue of eq.~(\ref{eq:C1w}) or~(\ref{eq:C2w}), the set of all integrable words of a given weight will thus be given by all linearly independent solutions of eq.~(\ref{eq:AX=0}), or in other words by the \emph{right nullspace} of the matrix $A$.

A systematic procedure for computing the nullspace is Gaussian elimination, whereby $A$ is brought into a column echelon form $H$ by a transformation $U$,
\begin{equation}
A\cdot U=H\,,
\end{equation}
\begin{equation}
U=(\,\underbrace{U_1}_r|N)\,,\quad H=(\,\underbrace{H_1}_r|\mathbf{0})\,.
\end{equation}
In the last line we have written out the two matrices in block matrix form, where $r$ denotes the rank of $A$, and the first nonzero element at each column of the invertible matrix $H_1$ is strictly below the corresponding element of the column at its left. Clearly, the submatrix $N$ will form a basis for the right nullspace of $A$.

Even though standard Gaussian elimination can be completed in a number of arithmetic operations that depends polynomially on the size of the system, a major complication that arises when applying it to matrices with exactly represented rational entries like $A$ is \emph{intermediate expression swell}: Generically, the size of the entries (in bits) doubles at each step, so that each operation takes longer and longer time, eventually leading to runtimes (and intermediate storage required) depending exponentially on the size of the system. A review of these well-established computer algebra results may be found in~\cite{Kauers:2008zz}.

The key idea for avoiding this complication is to transform our matrix from rational to integer, for which there exist fraction-free variants of Gaussian elimination that bound the size of intermediate expressions by virtue of Hadamard's inequality, see~\cite{Storjohann} and references therein. Fortunately, there already exists an efficient \texttt{C} library implementation of such a variant, the \texttt{Integer Matrix Library} (\texttt{IML})~\cite{Chen:2005:BBC:1073884.1073899}. This implementation also builds on the use of modular arithmetic to further improve the size of intermediate computations. Finally, it reduces row and column operations to matrix multiplications, which can be done very fast with the help of other well-known algorithms, for example~\cite{Strassen}.

First starting with the transformation of $A$ to an integer matrix $A^\prime$, we have found that a minimal increase in the size of its entries can be achieved by dividing each column of $A$ with the greatest common divisor (GCD) of all its nonzero elements (as opposed to doing this for the rows or even worse for the entire matrix). In fact, in this way we may also track down free variables which don't appear at all in the equations, as their columns will have zero GCD. These will correspond to the simplest nullspace vectors, which we can immediately construct and remove from the linear system, in order to reduce its size. If $D$ is the diagonal matrix whose diagonal elements are the inverses of the aforementioned GCDs, then $A,A^\prime$ will be related by $A^\prime=A\cdot D$, and we may obtain the nullspace of the former from the one of the latter,
\begin{equation}
A^\prime \cdot N^\prime=0=A\cdot N\quad \Rightarrow \quad N=D\cdot N^\prime\,.
\end{equation}
(Alternatively we may absorb the transformation into a rescaling of the unknown coefficients, $X^\prime=D\cdot X$.)

Once we have produced $A^\prime$ in this manner, we feed it as input into a custom \texttt{C} programme using the function \texttt{nullspaceLong} of the aforementioned \texttt{IML} library, which is optimised for matrices whose elements have absolute values smaller than $2^{31}$ (like the ones we have encountered), and computes a nullspace $N^\prime$ with integer entries. Particularly for the case of the weight 6 hexagon functions obeying last-entry conditions, we also found advantageous to apply this procedure not to $A^\prime$ directly, but to its Gram matrix $A^{\prime T}\cdot A^{\prime}$, exploiting the fact that the two matrices have the same nullspace,
\begin{equation}
A^{T}\cdot A\cdot X=0\Rightarrow X^T\cdot A^{T}\cdot A\cdot X=|A\cdot X|^2=0\Rightarrow A\cdot X=0\,.
\end{equation}
In this manner we traded a $63557 \times 15979$ matrix with a much smaller square $15979$ matrix (albeit with larger entries), and in fact with slightly fewer nonzero entries (corresponding to a fill-in of $1.6\%$ and $6.2\%$ approximately).

Finally it is worth mentioning that after obtaining $N$, whose columns span a basis of solutions for our linear system~(\ref{eq:AX=0}), we can further simplify this basis with the help of the Lenstra-Lenstra-Lov\'asz (LLL) algorithm, see~\cite{StorjohannLLL} for a more recent, improved version. The latter, which has also found applications in the computation of anomalous dimensions in SYM and QCD (see~\cite{Velizhanin:2014fua} and references therein), is an algorithm for finding a short, nearly orthogonal basis for a $d$-dimensional integer lattice embedded in $m$-dimensional space, $d\le m$. The integer matrix $N$ has precisely the form of such a lattice, where $d$ is the dimension of the nullspace, and $m$ the number of components of its vectors. In addition, it is evident from eq.~(\ref{eq:AX=0}) that any rescaling of the columns of $N$ will also be a nullspace basis. We can thus simplify our basis further by repeating a cycle of division of its vectors by the GCD of their nonzero elements, followed by an LLL reduction, until a new cycle leaves the basis unchanged. The final set of solutions to the integrability constraints has up to 3 times fewer nonzero coefficients than the initial set, leading to considerably shorter expressions for the corresponding integrable words~(\ref{eq:C1w}) or~(\ref{eq:C2w}).

\section{Heptagon Symbols and Their Properties}\label{sec:heptagonproperties}

Table~1 summarises the results of our partial analysis of the space of heptagon symbols through weight 6 (the question marks in the table indicate numbers that we have not yet explicitly determined). We remind the reader that in this paper we are working only at the level of symbols and that the counting of dimensions of spaces of functions obeying various constraints should be taken with this in mind. Following the conventions of subsection~\ref{subsec:heptagonfunctions}, in this section we continue to highlight this point by referring to {\bf heptagon symbols} (or {\bf hexagon symbols}) instead of the more cumbersome ``symbols of heptagon functions''.

We now discuss the results of Table~1 in detail, beginning with the first three lines which contain, perhaps, no great qualitative surprises.

\subsection{Collinear Limits of Heptagon Symbols}

The first line indicates the total number of heptagon symbols of a given weight, which we have already mentioned in section~\ref{subsec:heptagonfunctions}. The second line indicates the number of linear combinations of these which are finite in the collinear limit and independent of the ``bad'' letters shown in eq.~(\ref{eq:badletters}). Many linear combinations are not only well-defined, but actually vanish in the $7 \parallel 6$ collinear limit; the number of these is indicated on the third line.

For comparison with the hexagon bootstrap programme we include the analogous results for $n=6$ in Table~2. Here there is no distinction between the cases considered separately on lines 2 and 3 of Table~1: if the $6 \parallel 5$ collinear limit of a hexagon symbol is well-defined, then it necessarily vanishes in the limit, as there are no symbols for $n=5$.

\renewcommand{\arraystretch}{1.25}
\begin{table}[!ht]
\begin{center}
\begin{tabular}{|l|>{\hfill}p{.83cm}|>{\hfill}p{.83cm}|>{\hfill}p{.83cm}|>{\hfill}p{.83cm}|>{\hfill}p{.83cm}|>{\hfill}p{.83cm}|}
\hline\hline
\multicolumn{1}{|c|}{$~$ \hfill Weight $k=$}
&\multicolumn{1}{c|}{$1$}
&\multicolumn{1}{c|}{$2$}
&\multicolumn{1}{c|}{$3$}
&\multicolumn{1}{c|}{$4$}
&\multicolumn{1}{c|}{$5$}
&\multicolumn{1}{c|}{$6$}\\
\hline\hline
Number of heptagon symbols & 7 & 42 & 237 & 1288 & 6763 & ? \\
\hline
well-defined in the $7 \parallel 6$ limit & 3 & 15 & 98 & 646 & ? & ? \\
\hline
which vanish in the $7 \parallel 6$ limit & 0 & 6 & 72 & 572 & ? & ? \\
\hline
well-defined for all $i{+}1 \parallel i$ & 0 & 0 & 0 & 1 & ? & ? \\
\hline
with MHV last entries & 0 & 1 & 0 & 2 & 1 & 4 \\
\hline
with both of the previous two & 0 & 0 & 0 & 1 & 0 & 1 \\
\hline\hline
\end{tabular}
\caption{\label{fig:heptagon_functions} Heptagon symbols and their properties.}
\end{center}
\end{table}

\renewcommand{\arraystretch}{1.25}
\begin{table}[!ht]
\begin{center}
\begin{tabular}{|l|>{\hfill}p{.83cm}|>{\hfill}p{.83cm}|>{\hfill}p{.83cm}|>{\hfill}p{.83cm}|>{\hfill}p{.83cm}|>{\hfill}p{.83cm}|}
\hline\hline
\multicolumn{1}{|c|}{$~$ \hfill Weight $k=$}
&\multicolumn{1}{c|}{$1$}
&\multicolumn{1}{c|}{$2$}
&\multicolumn{1}{c|}{$3$}
&\multicolumn{1}{c|}{$4$}
&\multicolumn{1}{c|}{$5$}
&\multicolumn{1}{c|}{$6$}\\
\hline\hline
Number of hexagon symbols & 3 & 9 & 26 & 75 & 218 & 643 \\
\hline
well-defined (hence vanish) in the $6 \parallel 5$ limit & 0 & 2 & 11 & 44 & 155 & 516 \\
\hline
well-defined (hence vanish) for all $i{+}1 \parallel i$ & 0 & 0 & 2 & 12 & 68 & 307 \\
\hline
with MHV last entries & 0 & 3 & 7 & 21 & 62 & 188 \\
\hline
with both of the previous two & 0 & 0 & 1 & 4 & 14 & 59 \\
\hline\hline
\end{tabular}
\caption{\label{fig:hexagon_functions} Hexagon symbols and their properties.}
\end{center}
\end{table}

The collinear limit ties the two tables together in an interesting way, because the $7 \parallel 6$ collinear limit of a heptagon symbol must be a hexagon symbol, whenever the limit is well-defined. Of course, by taking collinear limits of all possible heptagon symbol we cannot possibly find more hexagon symbols than actually exist. This criterion partially explains the third line of Table~1. For example, at weights 1, 2, 3 we see by subtracting the third line from the second that there are 3, 9, 26 linearly independent hexagon symbols which can be obtained as collinear limits of heptagon symbols. These numbers match the top line of Table~2. So for weight $\le 3$ we conclude that the space of all hexagon symbols is spanned by the collection of (well-defined) collinear limits of heptagon symbols.

Curiously this pattern breaks down at weight 4. Table~1 indicates that taking the $7 \parallel 6$ collinear limit of heptagon symbols generates $646 - 572 = 74$ linearly independent hexagon symbols, but Table~2 indicates that there exist 75 hexagon symbols. Therefore, there is a weight-4 hexagon symbol which is not the collinear limit of any heptagon symbol!

\subsection{Symbols of Uniqueness: MHV Heptagons at 2 and 3 Loops}
\label{subsec:unique}

The real surprises in Table~1 lie in the last three lines, which stand out when compared to the last three lines of Table~2. Although the total number of heptagon symbols at a given weight is much greater (asymptotically exponentially) than the number of hexagon symbols at the same weight, the entries on the last three lines of the heptagon table are small compared to the corresponding entries in the hexagon table. The discovery of this surprising fact is the unexpected ``miracle'' of our work.

Let us begin with the fourth line of Table~1. A heptagon symbol may have a perfectly well-defined collinear limit as $7 \parallel 6$ while being divergent, or just ill-defined, in a different collinear limit, $4 \parallel 3$, say. (Note that nowhere in the two tables have we imposed cyclic symmetry.) MHV remainder functions must be finite and well-defined in all $i{+}1 \parallel i$ simple collinear limits. The number of heptagon symbols satisfying this criterion is indicated on the fourth line of Table~1. There are no such symbols for weight less than 4, and precisely one such symbol at weight 4. Obviously, that symbol must be the symbol of the two-loop seven-particle MHV remainder function $R_7^{(2)}$! To recap:

\begin{framed}
The symbol of the two-loop seven-particle MHV remainder function $R_7^{(2)}$ is the only weight-4 heptagon symbol which is well-defined in all $i{+}1 \parallel i$ collinear limits.
\end{framed}

Let us emphasise that it is not necessary to assume dihedral symmetry, parity symmetry, or the last-entry condition. Nor is it necessary to use the expected collinear limit $R_6^{(2)}$ as an input to fix some remaining ambiguity (except for the overall multiplicative normalisation). All of these properties are automatically satisfied by the unique function described in the above box. Of course, we have checked that the symbol of the function obtained in this manner via the bootstrap programme indeed is proportional to the known symbol of $R_7^{(2)}$ found in~\cite{CaronHuot:2011ky}.

It would be extremely interesting to see if this criterion continues to hold at weight 6, i.e.~to see whether the question mark in the last column of the fourth line of Table~1 is also 1, but we have not yet completed this calculation. Nevertheless we note that this criterion certainly could not work at arbitrary loop order; for example at weight 8 the square of $R_7^{(2)}$ and the four-loop seven-particle MHV remainder function $R_7^{(4)}$ are both well-defined in all simple collinear limits, and are distinct.

Let us now turn to the last two lines of Table~1, where we impose the last-entry condition appropriate for MHV amplitudes, as discussed in section~\ref{subsec:qbar}. In contrast to the general heptagon problem, where the complexity of the linear systems involved has forced us to leave some questions marks in the table, when we impose the last-entry condition the size of the linear systems becomes ``small'' enough that we have succeeded in a full classification through weight 6. This is certainly not to say that the calculation was easy---as explained in section~\ref{sec:methods}, determining the number ``4'' in the last column of Table~1 required finding the nullspace of a linear system with over half a million variables.

The number of heptagon symbols satisfying the last-entry condition is shown in the sixth line of Table~1. Let us note right away that none of the numbers are multiples of 7, hence all of these functions are necessarily cyclically invariant, even though this was not an input to the calculation. Also it turns out (this is trivial at weights 2 and 5, where there is a single symbol, and is easily checked at weights 4 and 6) that they are all invariant under the full dihedral group, as well as under the parity operation shown in eq.~(\ref{eq:parity}). Again none of these discrete symmetries were imposed going into the calculation.

At weight 2 we find there is a unique heptagon symbol satisfying the last-entry condition. The corresponding heptagon function is written explicitly, and discussed in more detail, in the following section. At weight 4 there are two functions: the square of the weight-2 function, and the two-loop MHV remainder function $R_7^{(2)}$. As we have already seen, the latter is the only one which is well-defined in all collinear limits. At weight 5 there is again a unique symbol satisfying the last-entry condition. This weight-5 symbol, like the weight-2 symbol, is not well-defined in the collinear limit, so these two symbols have no role to play in connection with MHV scattering amplitudes.

Let us now focus again on the surprising ``4'' in the last column of Table~1. From our discussion so far we already know that there must be at least two weight-6 heptagon functions satisfying the last-entry condition: the cube of the weight-2 function discussed above, and the product of the weight-2 function with $R_7^{(2)}$. The surprise is that, in addition to the symbols of these two functions, we find only two irreducible symbols at weight 6. We find that there is a unique linear combination of these four symbols which is finite in the collinear limit (in this case it happens that it is sufficient to consider only the $7 \parallel 6$ collinear limit since, as mentioned above, the symbols turn out to all be cyclically invariant anyway). We have checked that the collinear limit matches perfectly (up to an overall factor, which is not fixed by the bootstrap) the known symbol of the three-loop MHV hexagon~\cite{Dixon:2011pw,CaronHuot:2011kk}. Therefore:

\begin{framed}
The symbol of the three-loop seven-particle MHV remainder function $R_7^{(3)}$ is the only weight-6 heptagon symbol which satisfies the last-entry condition and which is finite in the $7 \parallel 6$ collinear limit.
\end{framed}

The only ambiguity which will be left when passing from the symbol of $R_7^{(3)}$ to an actual function is the addition of a rational linear combination of $\zeta_2 R_7^{(2)}$, $\zeta_3^2$ and $\zeta_6$. The collinear limit will fix all three coefficients (as well as the overall normalisation of $R_7^{(3)}$) uniquely.

Again we emphasise that the above conclusion does not rely on assuming that any of the discrete symmetries are satisfied; they all emerge as ``accidental'' (if there is such a thing in SYM theory) properties of the unique solution. Moreover, and even more surprisingly, the unique solution emerges without any free parameters which need to be tuned in order to match the correct value of the three-loop MHV hexagon in the collinear limit, let alone to match various terms in the Regge limit and/or OPE expansion around the collinear limit.

\section{Speculations: The $n$-gon Bootstrap at Weight 2}\label{sec:speculations}

Our results were completely surprising. Based on the hexagon bootstrap programme, we expected that even after imposing all discrete symmetries, there would likely be hundreds of free parameters in our heptagon ansatz which would need to be fit by comparison to various data in the literature.

The fact that none of this turned out to be necessary, and that the heptagon bootstrap turned out, in this sense, to be more powerful than the hexagon bootstrap, requires explanation. It is, after all, a basic tenet of amplitudeology that ``accidents do not happen,'' especially in SYM theory.

Unfortunately we have only very little to offer at this time. In this section we make a few meager observations at weight 2, where it is simple to tabulate and explicitly analyze the relevant function spaces. Our observations here admittedly shed only a little light on the situation at higher weight, but perhaps they serve as a useful starting point.

Let us define the cross-ratio
\begin{equation}
u_1 = \frac{\vev{1256} \vev{2345}}{\vev{1245} \vev{2356}} = \frac{a_{17}}{a_{13} a_{14}}
\end{equation}
with six other $u_i$ defined cyclically (sometimes $u_1$ is called $u_{14}$ in the literature). According to the sixth line of Table~1, there is a unique weight-2 heptagon function satisfying the last-entry condition. This function is
\begin{equation}
\label{eq:weight2}
\sum_{i=1}^7  \Li_2(1 - 1/u_i)
+ \frac{1}{2} \log u_i \log \frac{u_{i+2} u_{i-2}}{u_{i+3} u_i u_{i-3}}\,.
\end{equation}
Let us contrast this to the situation at $n=6$, where there are three functions
\begin{equation}
\label{eq:degreetwofunction}
\Li_2(1-1/u)\,, \qquad \Li_2(1-1/v)\,, \qquad \Li_2(1-1/w)
\end{equation}
which separately satisfy the last-entry condition.

Why does this happen? The functions shown in eq.~(\ref{eq:degreetwofunction}) exist because of the identity
\begin{equation}
\label{eq:identity6}
\frac{1-u}{u} =
\frac{\vev{1356} \vev{2346}}{\vev{1236} \vev{3456}}
\end{equation}
(and two cyclic images). Note that all of the brackets appearing on the right-hand side are of the form $\vev{i\,j{-}1\,j\,j{+}1}$. Hence all three of ${(1-u)/u,(1-v)/v,(1-w)/w}$ are valid MHV last entries.

Is there an analogue to the identity~(\ref{eq:identity6}) for $n=7$? That is, does there exist an identity which allows products of $u_i$'s and $1-u_i$'s to be rewritten only in terms of $\vev{i\,j{-}1\,j\,j{+}1}$'s? It is simple to check that there are precisely seven such identities:
\begin{equation}
\label{eq:identity7}
\frac{1-u_1}{u_1} \frac{1-u_7}{u_7} \frac{1}{1-u_4}
= \frac{\vev{1235} \vev{1247} \vev{1345} \vev{2456}}{\vev{1234} \vev{1257} \vev{1456} \vev{2345}}
\end{equation}
and its cyclic permutations. If it were not for the factor $1/(1-u_4)$ on the left, then the functions $\Li_2(1-1/u_i) + \Li_2(1-1/u_{i+1})$ would satisfy both the first- and last-entry conditions for all $i$. Instead, only the particular linear combination shown in eq.~(\ref{eq:weight2}) is allowed.

It is straightforward to extend this analysis to higher $n$. The total number of weight-2 $n$-gon functions grows very rapidly (in fact, as $\mathcal{O}(n^4)$) with $n$. How many of those functions satisfy the last-entry condition? Obviously, it is to be expected that there should be very strong interplay between the first- and last-entry conditions at weight 2, but we find an unexpectedly strong result:

We find that there are precisely 3 $n$-gon functions at weight 2 satisfying the last-entry condition when $n=6$ or when $n$ is a multiple of four. For all other $n$, we find that there is only one such function!

As we go to higher weight we might expect the interplay between the first- and last-entry conditions to become less constraining. This expectation may or may not turn out to be true asymptotically at large weight, but Tables~1 and~2 indicate little weakening of this interplay for $k$ even as high as 6. Clearly it would be interesting to map out the space of $n$-gon functions at higher weight.

\section{Discussion}

We have found that the heptagon bootstrap for computing (symbols of) seven-point MHV amplitudes in SYM is unreasonably effective in comparison with the hexagon bootstrap, at least through three loops. In particular, the three-loop heptagon remainder function is the unique weight-6 heptagon function which satisfies the last-entry condition and which is finite in the $7 \parallel 6$ collinear limit. Evidently the conceptually simplest way of computing the three-loop hexagon remainder function is, somewhat perversely, to first compute the heptagon remainder and take its collinear limit.

Naturally, it would be very interesting to further explore the power of the heptagon bootstrap at higher loops or by relaxing the last-entry condition to those appropriate for NMHV amplitudes. It would also be interesting to explore the $n$-gon bootstrap for higher $n$. Our analysis at weight-2 in section~\ref{sec:speculations} suggests that the first- and last-entry conditions are much tighter in combination than each is individually. It would be important to understand whether this is an accident at weight-2 (and whether the success of our heptagon bootstrap was similarly accidental), or whether there is some fundamental feature of the structure of $n$-gon functions which currently evades our understanding.

The cases $n=6,7$ are special because we believe that we know the appropriate symbol alphabets for amplitudes (both MHV and non-MHV) to all loop order, based on the fact that the associated cluster algebras have finitely many $\mathcal{A}$-coordinates. However starting at $n=8$ their number is infinite, so there is the possibility that new, more exotic symbol letters could start appearing at each loop order (or even when we go from MHV to non-MHV at a given loop order). Anything we could learn about the pattern of symbol letters which appear at higher $n$ and at higher weight would be very valuable.

In section~\ref{sec:heptagonproperties} we found an indication that thinking about the collinear limits of $n$-gon functions may lead to a class of previously underappreciated constraints. Specifically we found that there exists a hexagon function at weight 4 which is not the collinear limit of any heptagon function. Similarly, it is natural to expect that there may be heptagon functions which are not the collinear limit of any octagon functions, that there are hexagon functions which are not the double-collinear limit of any octagon function, etc. In this way we see that the consistency of collinear limits places an entire infinite tower of potentially very powerful constraints on the bootstrap. Along these lines, it has recently been shown~\cite{Golden:2014pua} that the collinear limit, together with dihedral symmetry and the first- and last-entry conditions, uniquely fixes the two-loop $n$-point MHV amplitude modulo classical polylogarithm functions for all $n$. The results of this paper suggest that even full symbols, if not full functions (which we have not addressed), may be surprisingly accessible via the $n$-gon bootstrap.

Finally, it is of course important to construct a complete functional representation for $R_7^{(3)}$. This would require first constructing a heptagon function of weight 6 with the correct symbol, obeying the differential $\bar Q$ constraint and finite in the collinear limit. After this there will be beyond-the-symbol ambiguities corresponding to the addition of a numerical constant (in particular a linear combination of $\zeta_3^2$ and $\zeta_6$) as well as a term proportional to $\zeta_2 R_7^{(2)}$. Both of these ambiguities will be uniquely fixed by the simple collinear limit. Having an explicit functional form for $R_7^{(3)}$ would not only allow for detailed checks against the available predictions for its behaviour in the collinear~\cite{Basso:2013aha,Basso:2014nra} and multi-Regge limits~\cite{Bartels:2011ge,Bartels:2013jna,Bartels:2014jya}, but would also shed light on yet unknown key quantities in these approaches. These include multi-particle scalar and fermion pentagon transitions, or higher BFKL eigenvalues, impact factors and central emission vertices. For example it would be great if our result could guide the generalisation of the all-loop formulas for the hexagon in the multi-Regge limit~\cite{Basso:2014pla}, to the heptagon. The continuation of this programme for $n=8$ will have an even more interesting interplay with the BFKL approach, where a new bound state of three reggeised gluons first appears, and could moreover push forward our knowledge of the strong coupling behaviour of the amplitudes~\cite{Bartels:2014ppa,Bartels:2014mka}.

\section*{Acknowledgements}

We thank B.~Basso, S.~Caron-Huot, L.~Dixon, A.~Sever and P.~Vieira for comments on the draft, and acknowledge having benefitted from stimulating discussions with J.~Golden, E.~Sokatchev and A.~Volovich. The work of GP was supported by the French National Agency for Research (ANR) under contract StrongInt (BLANC-SIMI-4-2011). The work of MS was supported by the US Department of Energy under contract DE-SC0010010. MS is also grateful to the CERN theory group for their hospitality and support.

\bibliographystyle{JHEP}
\bibliography{heptagon}

\end{document}